\documentclass[modern]{aastex63}

\received{xxxx}
\revised{yyyy}
\accepted{zzzz}
\shorttitle{Radial velocity performance of ESPRESSO}
\shortauthors{Netto et al.}
\graphicspath{{./}{figures/}}
\begin{document}

\title{Radial velocity precision of ESPRESSO through the analysis of the solar twin HIP 11915}

\correspondingauthor{Yuri Netto}
\email{yurinetto@usp.br}

\author{Yuri Netto}
\affiliation{Universidade de São Paulo, Instituto de Astronomia, Geofísica e Ciências Atmosféricas (IAG), Departamento de Astronomia, Rua do Matão 1226, Cidade Universitária, 05508-900, SP, Brazil}

\author{Diego Lorenzo-Oliveira}
\affiliation{Universidade de São Paulo, Instituto de Astronomia, Geofísica e Ciências Atmosféricas (IAG), Departamento de Astronomia, Rua do Matão 1226, Cidade Universitária, 05508-900, SP, Brazil}

\author{Jorge Meléndez}
\affiliation{Universidade de São Paulo, Instituto de Astronomia, Geofísica e Ciências Atmosféricas (IAG), Departamento de Astronomia, Rua do Matão 1226, Cidade Universitária, 05508-900, SP, Brazil}

\author{Jhon Yana Galarza}
\affiliation{Universidade de São Paulo, Instituto de Astronomia, Geofísica e Ciências Atmosféricas (IAG), Departamento de Astronomia, Rua do Matão 1226, Cidade Universitária, 05508-900, SP, Brazil}

\author{Raphaëlle D. Haywood}
\affiliation{Astrophysics Group, University of Exeter, Exeter EX4 4QL, UK}

\author{Lorenzo Spina}
\affiliation{INAF Osservatorio Astronomico di Padova, vicolo dell’Osservatorio 5, 35122, Padova, Italy}

\author{Leonardo A. dos Santos}
\affiliation{Observatoire Astronomique de l’Universit\'{e} de Gen\`{e}ve, Chemin Pegasi 51, 1290 Versoix, Switzerland}



\begin{abstract}

Different stellar phenomena affect radial velocities (RVs), causing variations large enough to make it difficult to identify planet signals from the stellar variability. RV variations caused by stellar oscillations and granulation can be reduced through some methods, but the impact of rotationally modulated magnetic activity on RV, due to stellar active regions is harder to correct. New instrumentation promises an improvement in precision of one order of magnitude, from about 1 m/s, to about 10 cm/s. In this context, we report our first results from 24 spectroscopic ESPRESSO/VLT observations of the solar twin star HIP 11915, spread over 60 nights. We used a Gaussian Process approach and found for HIP 11915 a RV residual RMS scatter of about 20 cm s$^{-1}$, representing an upper limit for the performance of ESPRESSO.

\end{abstract}



%%%%%%%%%%%%%%%%%%%%%%%%%%%%%%%%%%%%%%%%%%%%%%%%%%%%%%%%%%%%%%%%%%%%%%%%%%%%%%%%
\section{Introduction}

The radial velocity (RV) method has been used to discover exoplanets, or to confirm exoplanets detected by the transit method and estimate their mass. Earth analogs orbiting the habitable zone of a Sun-like star, induce a RV signal on the order of 10 cm s$^{-1}$ \citep[e.g.,][]{2020Langellier}. However, the measured RV variations are not entirely due to planets; stellar activity, oscillations and granulation also cause changes in RV \citep[e.g.,][]{2016Fischer}. The effect of stellar oscillations and granulation on RV can be reduced by adopting exposure times tailored to the star's spectral type \citep{2019Chaplin}, or modelling through magnetohydrodynamical simulations \citep{2016Cegla}. Yet, the impact of rotationally modulated stellar activity is hard to predict, as stellar activity cycles are not strictly periodic and have complex shapes. Futhermore, the presence of active regions on the stellar surface can induce RV signals, which can sometimes be mis-interpreted as planet signals \citep[e.g.][]{2010Figueira, 2014Haywood, 2018Diaz}.

Thus, a better understanding and characterisation of the impact of stellar activity on RV is crucial for the progress in our ability to detect exoplanets \citep{2020Blackwood}. After the discovery of the first exoplanet around solar-type stars \citep{1995MayorQueloz}, \citet{1997SaarDonahue} discussed the possible impact of apparent changes in RV due to the effects of stellar magnetic activity. Further, \citet{2002Hatzes} and \citet{2007Desort} also presented characterisation of the likely effects of stellar active regions. More recently the interest has been renewed \citep[e.g.,][]{2020Haywood, 2018Bauer, 2015Borgniet, 2015Korhonen, 2014Dumusque}, as planet searches are focusing in small exoplanets, for which the stellar activity imposes serious limitations in their detection.

Observations and models show the dependence of the RV-activity on the activity cycle phase \citep[e.g.,][]{2015Borgniet, 2015Korhonen}, being the lowest at the activity cycle minimum. There have been important advances in the treatment of star-induced RV variability, through reconstructing the geometry of active regions \citep[e.g,][]{2014Dumusque} or semi empirically through a Gaussian process (GP) fit \citep[e.g,][]{2014Haywood}. GP regression \citep{2015Rajpaul, 2012Roberts} has become one of the most successful tools in the analysis of stellar activity in RV time series \citep{2017Dumusque}.

The aim of this work is to test the accuracy of the RV uncertainties estimated by the ESPRESSO data reduction pipeline, reported to be at a precision level of 10 cm s$^{-1}$ \citep{2020Pepe}. For this, we analyse RV observations of the solar twin HP 11915, which is currently in the minimum of its magnetic activity cycle; we account for rotationally modulated stellar activity using GP regression. Details regarding the data are in Section~\ref{sec:observation}. In Section~\ref{sec:rv} we present the method to analyse the RV. We discuss our results in Section~\ref{sec:conclusion}.

%%%%%%%%%%%%%%%%%%%%%%%%%%%%%%%%%%%%%%%%%%%%%%%%%%%%%%%%%%%%%%%%%%%%%%%%%%%%%%%%
\section{Observations and data reductions} \label{sec:observation}

The new high-resolution spectrograph ESPRESSO \citep[\textit{Echelle SPectrograph for Rocky Exoplanets and Stable Spectroscopic Observations},][]{2014Pepe, 2020Pepe} of ESO’s Very-Large Telescope (VLT) started operations at the ESO Paranal observatory in September 2018. It is designed and built to detect and characterise Earth analogs within the habitable zones of their host stars, reaching RV precision of 10 cm s$^{-1}$. More details regarding ESPRESSO may be found in the general description of the ESPRESSO instrument, reported on the actual on-sky performance described by \cite{2020Pepe}.

The solar twin star HIP 11915 was selected from our former HARPS planet survey around solar twins, from which a Jupiter-twin planet was found \citep{2015Bedell}. The star has been carefully characterised in our previous works \citep{2021Galarza, 2018Spina, 2016dosSantos}, with improved Ca II H$\&$K activity indices derived by \citet{2018LorenzoOliveira}. In Table~\ref{tab:objects} we summarise the main fundamental parameters derived for HIP 11915. Within the ESO program 0102.C-0523 (PI: Jorge Mel\'{e}ndez), we obtained a total of 24 observations of HIP 11915 from November 2018 to January 2019. Measurements were taken in ESPRESSO's High Resolution 1-UT (HR) mode to reach a high resolving power ($R = \lambda/\Delta \lambda = 140 \ 000$). We reduced the spectra using the 1.3.2\footnote{https://www.eso.org/sci/software/esoreflex/} version of the ESOReflex environment \citep{2013Freudling}. 
The data reduction includes the standard procedures such as corrections for bias, flat-field and background light, wavelength calibration, extraction of the spectrum, merging of the Echelle orders, and barycentric and instrumental drift corrections. The resulting spectra are given in counts and calibrated in flux. The wavelength calibration is performed combining a Thorium-Argon (Th-Ar) hollow-cathode lamp and a white-light illuminated Fabry-Pérot. The former provides wavelength accuracy, and the latter ensures wavelength precision. The pipeline also provides a cross-correlation function (CCF), computed with respect to a binary template mask. Then, the RV is obtained from a Gaussian fit to the CCF. For each exposure, we extracted the RV and activity indicators from the CCF (Full width at half maximum, FWHM, and CCF CONTRAST). The activity index S is measured from the extracted spectra, following prescriptions given in \citet{2004Wright}. The contrast of the CCF, expressed as a percentage, is the relative depth of the CCF at its central wavelength, used to measure temporal changes \citep{2020Lafarga}. \citet{2019Maldonado} found significant correlations between the contrast with the main optical activity indicators for the Sun-as-a-star observations. 

\begin{deluxetable*}{ccc}
\tablenum{1}
\tablecaption{Fundamental parameters for HIP 11915.\label{tab:objects}}
\tablewidth{0pt}
\tablehead{
\colhead{Parameter} & \colhead{Value} & \colhead{Reference}
}
%\decimalcolnumbers
\startdata
Spectral Type                         & G5V                  & {[}1{]}   \\
V(mag)                                & 8.615                & {[}2{]}   \\
$T_{\rm eff}$(K)                      & 5773 $\pm$ 2         & {[}3{]}   \\
{[}Fe/H{]}                            & $-$0.057 $\pm$ 0.003 & {[}3{]}   \\
$\log~g$                              & 4.470 $\pm$ 0.008    & {[}3{]}   \\
$\log R^\prime_{\rm HK}(T_{\rm eff})$ & $-$4.923 $\pm$ 0.028 & {[}4{]}   \\
Age (Gyr)                             & 3.87 $\pm$ 0.39      & {[}3{]}   \\
Mass (M{$_\odot$})                    & 0.991 $\pm$ 0.003    & {[}3{]}   \\
$P_{\rm rot}/\sin(i)$(d)              & 42.0 $\pm$ 8.6       & {[}5{]}   \\
Number of observations                & 24                   &   
\enddata
\tablecomments{[1] \citet{1988Houk}, [2] \citet{2012Ramirez}, [3] \citet{2021Galarza}, [4] \citet{2018LorenzoOliveira}, [5] \citet{2019LorenzoOliveira}}
\end{deluxetable*}

%%%%%%%%%%%%%%%%%%%%%%%%%%%%%%%%%%%%%%%%%%%%%%%%%%%%%%%%%%%%%%%%%%%%%%%%%%%%%%%%
\section{RV performance through GP analysis} \label{sec:rv}

The star-induced RV variability is difficult to model deterministically, and has motivated the use of GP regression \citep{2014Haywood, 2015Rajpaul}. GP is a powerful statistical technique, in which instrinsic stellar variability is treated as correlated noise described by a covariance functional form, while the hyper-parameters describe the physical phenomena to be modelled. A kernel function is chosen to model these covariances as a function between two measurements at a time \citep{2014Haywood, 2015Rajpaul, 2016Faria}.

\subsection{Gaussian Process Model}

For a given activity indicator $i$, we define a combination of covariance functions between two measurements at time $t$ and $t'$ of the observations, to build our quasi-periodic (QP) activity model:

\begin{equation}
    k(t,t')\equiv \mathcal{I}_{\rm const}+\mathcal{A}\exp\left(-\frac{||t-t'||}{2 \ell^2} -\Gamma\sin^2\left[\frac{\pi}{ P_{\rm rot}}||t-t'||\right] \right)+\sigma^2\delta_{t,t'},
\end{equation}

\noindent where $\mathcal{I}_{\rm const}$ gives a constant scale to match the observed mean activity level of the star, while $\mathcal{A}$ represents the amplitude of the rotation signal. The timescale of rising and decay of active regions is interpreted by $\ell$, the harmonic nature of the time series is represented by $\Gamma$, the white noise term is $\sigma^2\delta_{t,t'}$ and P$_{rot}$ is the rotational period, where P$_{rot}$ was initially determined by using generalised Lomb-Scargle periodograms \citep[GLS,][]{2009Zechmeister}. We use a log-normal prior distribution for the hyper-parameters (HP): $\mathcal{I}_{\rm const}$ ($\mu$ = $<$ i $>$, $\sigma$ = $\sigma_{<\text{i}>}$), $P_{\rm rot}$ ($\mu$ = $P_{\rm rot}$,$\sigma$ = 0.2 $\times$ P$_{rot, GLS}$), and ln $\Gamma$ ($\mu$ = $-$2.3, $\sigma$ = 1.4, as in \citealp{2018Angus}). We use the \textit{emcee} \citep{2013ForemanMackey} Python implementation of the affine-invariant Markov chain Monte Carlo method (MCMC) following \citet{2018Angus} to estimate the posterior distribution of all GP HP. In brief, we start the MCMC process with 64 walkers to sample the parameter space, and are initialised with an optimal solution obtained by maximum likelihood optimisation. Then, every 100 steps, we evaluate the convergence of chains and check its auto-correlation timescale ($\tau_{chain}$) and the consistency of walkers solutions through Gelman-Rubin statistics (\^R) \citep{1992Andrew}. We consider that the walkers have converged when $\tau_{chain}$ is less than 10 $\%$ of the total chain length, $\tau_{chain}$ is stable concerning the previous chain evaluation within 1$\%$, and \^R less than 1.03. Finally, we discard the initial iterations (3$\times \tau_{chain}$) and randomly re-sample 5000 samples to represent our estimate of posterior probability distribution of all GP hyper-parameters.

\subsection{Fitting the RV}

\begin{figure}
\centering
\includegraphics[height=4cm]{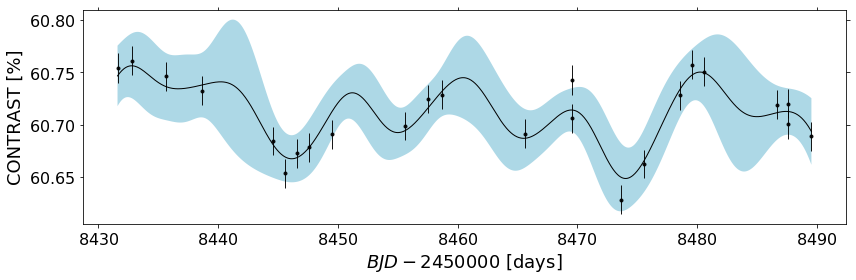}
\includegraphics[height=4cm]{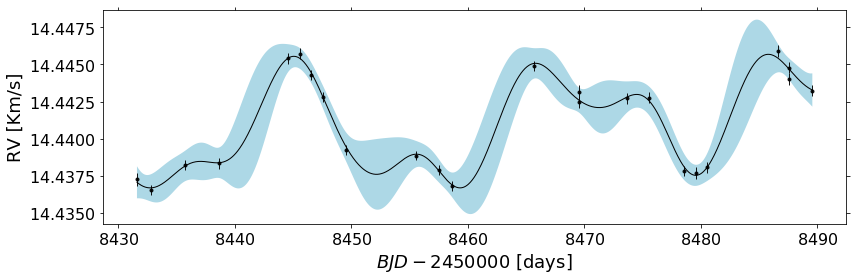}
\caption{GP regression fits to $\sim$ 60 days of CONTRAST (upper panel) and RV data (lower panel) of HIP 11915, both from ESPRESSO. Solid lines and shaded blue regions are the QP model and 2$\sigma$ predictions, respectively.
\label{fig:GPFIT_CONTRAST}}
\end{figure}

To model the RV of HIP 11915 we fit the activity indicator $i$ and the RV consecutively. We fit the RV using the same QP kernel function of equation 1 with the $\Gamma$, P$_{cycle}$ and $\ell$ fixed at the median value from the activity indicator $i$ fit. The GP regression fit to the CONTRAST and to the RV are shown in Fig. \ref{fig:GPFIT_CONTRAST}. The resulting HP estimates are displayed in the last column of Table \ref{tab:GP_MCMC} as the median and 16 $\%$ confidence interval of these uncorrelated samples. We also applied the same procedure to the FWHM and the $S$ Index and its resulting HP estimates for the RV are displayed in the bottom part of the Table \ref{tab:GP_MCMC}.

\begin{deluxetable*}{ccccc}
\tablenum{2}
\tablecaption{Summary of hyper-parameters, upper and lower bounds of the priors and its initial guesses used for Markov chain Monte Carlo (MCMC) sampling for the CONTRAST (CONT) and RV fits of ESPRESSO data. The last column shows the resulting median value of the MCMC samples and their corresponding error. \label{tab:GP_MCMC}}
\tablewidth{0pt}
\tablehead{
\colhead{} & \colhead{Fit HP} & \colhead{MCMC initial guess} & \colhead{Priors Bounds} & \colhead{Fit Value}
}
%\decimalcolnumbers
\startdata
        & $\mathcal{I}_{\rm CONT}$ [$\%$]  & mean(CONT)       & (60.628, 60.761) & 60.710 $^{+0.021}_{-0.023}$ \\
        & $\sigma_{\rm CONT}$ [$\%$]       & mean(CONT error) & (0.0022, 0.0226) & 0.0134 $^{+0.0043}_{-0.0033}$ \\
CONT HP & $\mathcal{A}_{\rm CONT}$ [$\%$]  & var(CONT)        & (0.034, 0.174) &  0.059 $^{+0.030}_{-0.016}$ \\
        & $\ell$ [d]                              & 30               & (2, 365) &   68.8 $^{+68.1}_{-27.6}$ \\
        & $\Gamma$                                & 0.1              & (-10, 2) &    2.5 $^{+1.9}_{-1.3}$ \\
        & P$_{rot}$ [d]                           & 24.03 (GLS)      & (15, 50) &   27.4 $^{+1.2}_{-7.9}$ \\
        & $\mathcal{I}_{\rm CONT, RV}$ [km s$^{-1}$] & mean(RV)         & (14.436, 14.446) & 14.441 $\pm$ 0.002 \\
RV HP   & $\sigma_{\rm CONT, RV}$ [km s$^{-1}$]      & mean(RV error)   & (10$^{-4}$, 5$\times10^{-3}$) & 0.0002 $\pm$ 0.0001\\
        & $\mathcal{A}_{\rm CONT, RV}$ [km s$^{-1}$] & var(RV)          & (0.0032, 0.0160) & 0.0039 $^{+0.0008}_{-0.0037}$\\
        \\
        \hline
        & $\mathcal{I}_{\rm FWHM, RV}$ [km s$^{-1}$] & mean(RV)         & (14.436, 14.446) & 14.441 $\pm$ 0.002 \\
RV HP   & $\sigma_{\rm FWHM, RV}$ [km s$^{-1}$]      & mean(RV error)   & (10$^{-4}$, 5$\times10^{-3}$) & 0.0002 $\pm$ 0.0001\\
        & $\mathcal{A}_{\rm FWHM, RV}$ [km s$^{-1}$] & var(RV)          & (0.0032, 0.0160) & 0.0037 $^{+0.0006}_{-0.0003}$\\
        \\
        & $\mathcal{I}_{\rm S Index, RV}$ [km s$^{-1}$] & mean(RV)         & (14.436, 14.446) & 14.441 $\pm$ 0.001 \\
RV HP   & $\sigma_{\rm S Index, RV}$ [km s$^{-1}$]      & mean(RV error)   & (10$^{-4}$, 5$\times10^{-3}$) & 0.0002 $\pm$ 0.0001\\
        & $\mathcal{A}_{\rm S Index, RV}$ [km s$^{-1}$] & var(RV)          & (0.0032, 0.0160) & 0.0037 $^{+0.0005}_{-0.0003}$  
\enddata
\end{deluxetable*}

\begin{figure}
\centering
\includegraphics[height=4cm]{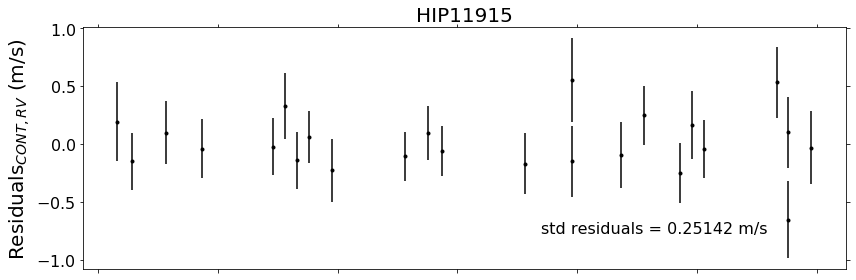}
\includegraphics[height=4cm]{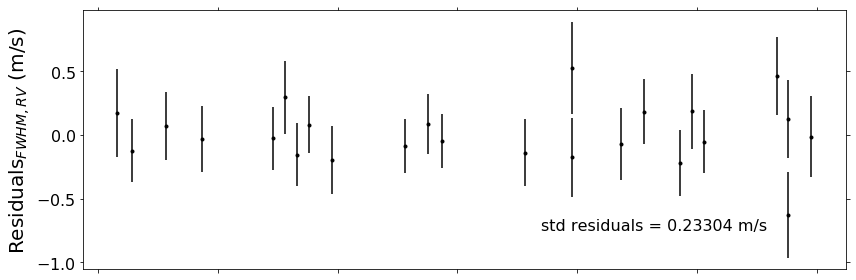}
\includegraphics[height=4cm]{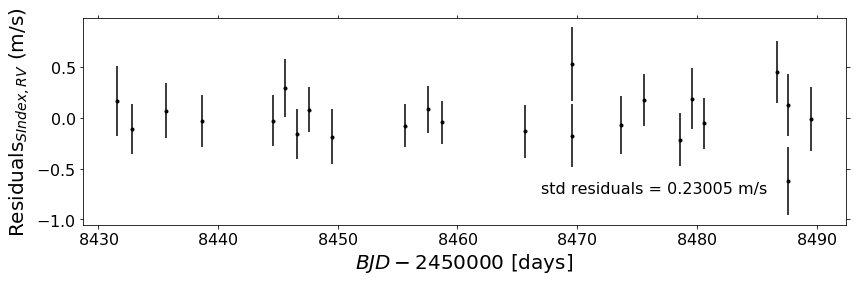}
\caption{Residuals in radial velocities of the GP fit for each activity indicator with its rms scatter. In each panel, the GP was trained on: \textbf{Top:} CONTRAST; \textbf{Middle:} FWHM; \textbf{Bottom:} $S$ Index.
\label{fig:Residuals}}
\end{figure}

The root mean square (RMS) of the residuals is 25 cm s$^{-1}$, 23 cm s$^{-1}$ and 23 cm s$^{-1}$ for the RMS res$_{\rm CONTRAST, RV}$, RMS res$_{\rm FWHM, RV}$ and RMS res$_{\rm S Index, RV}$, respectively. These values can be compared to the average RV uncertainty of 27cm s$^{-1}$. They are also in agreement with the posterior estimate for the white noise term parameters of 20 cm s$^{-1}$, yet at the level of few tens of cm s$^{-1}$ . In Fig. \ref{fig:Residuals}, we show all the residuals for each fit. The values reached for the RV precision are robust, since the GP was applied independently for each activity indicator. 

To test the stability of the values found for the total sample, we re-sampled 30 times, choosing the observations randomly, keeping 2/3 of the total sample. The results remained relatively stable, deviating +6 cm s$^{-1}$ from the original values, except for the FWHM which had the largest deviation, 11 cm s$^{-1}$. The HP also remained stable in tests, most of them resulting in the same value found for the total sample. The HP $\ell$ and $\Gamma$ present greater variation in results, from $\sim$ 60 days to $\sim$ 200 days and $\sim$ 0.5 to $\sim$ 2.5, respectively.
 
We also used the GP approach applied to the recently released ESO Phase 3 (pipeline version 2.2.1 or higher), to test if different pipelines to reduce the data interfere with the results. The results remained relatively stable, the root mean square (RMS) of the residuals is 24 cm s$^{-1}$, 41 cm s$^{-1}$ and 44 cm s$^{-1}$ for the RMS res$_{\rm CONTRAST, RV}$, RMS res$_{\rm FWHM, RV}$ and RMS res$_{\rm S Index, RV}$, respectively.

%%%%%%%%%%%%%%%%%%%%%%%%%%%%%%%%%%%%%%%%%%%%%%%%%%%%%%%%%%%%%%%%%%%%%%%%%%%%%%%%
\section{Conclusions} \label{sec:conclusion}

In this work we report results about the ESPRESSO RV precision. Using the solar twin HIP 11915 data ($\sim$ 60 days) from the ESPRESSO spectrograph at the VLT, we applied the Gaussian Process regression with a quasi-periodic kernel to obtain the residuals from the fits.
We found an average value of 24 cm s$^{-1}$ for the RMS of our residuals, that represents an upper limit for the performance of ESPRESSO for this range of observation. HIP 11915 is a solar twin star, observed during the minimum of its activity cycle, and the low activity level of the star has helped to evaluate the performance of ESPRESSO, demonstrating that the instrument can achieve low-to-mid 20 cm/s on quiet stars.

With long-term repeated measurements, it may be possible to improve the RV precision for the HIP 11915. This is a great target for the searches of Earth analogs, since it has a Jupiter twin \citep{2015Bedell} , has a chemical composition depleted in rocky-forming elements, similar to the Sun \citep{2021Galarza}, and a precision at the level of 20 cm s$^{-1}$ can be achieved.

%%%%%%%%%%%%%%%%%%%%%%%%%%%%%%%%%%%%%%%%%%%%%%%%%%%%%%%%%%%%%%%%%%%%%%%%%%%%%%%%
\acknowledgments

Y.N. and J.M. thanks support from FAPESP (2019/20319-8; 2018/04055-8). D.L.O. and J.M.thanks support from FAPESP (2016/20667-8; 2018/04055-8). J.Y.G. acknowledges the support from CNPq (142084/2017-4). LAdS acknowledges the support of NCCR PlanetS, the Swiss National Science Foundation, and the European Research Council under programme No 724427 (project {\sc Four Aces}).

\vspace{5mm}
\facility{\textbf{ESO}: VLT 8.2-meter Unit Telescopes, Echelle SPectrograph for Rocky Exoplanet and Stable Spectroscopic Observations (ESPRESSO).}





\bibliography{sample63}{}
\bibliographystyle{aasjournal}



\end{document}